\documentclass[pre,singlecolumn,amsmath,amssymb,longbibliography]{revtex4-1}
\usepackage{amsmath,amssymb,latexsym,epsfig,graphics,epsf,bm}
\usepackage{graphics,xcolor}
\usepackage{float}
\newcommand{\teq}{\tau_{\rm eq}}
\newcommand{\geqa}{\gamma_{\rm eq}}
\newcommand{\DUto}{\langle (\Delta U)^2\rangle}

\newcommand{\eq}[1]{Eq.~(\ref{#1})}
\newcommand{\Eq}[1]{Equation~(\ref{#1})}
\newcommand{\be}{\begin{equation}}
\newcommand{\ee}{\end{equation}}

\newcommand{\DU}{\Delta U}

\newcommand{\fig}[1]{Fig. ~\ref{#1}}
\newcommand{\Fig}[1]{Figure~\ref{#1}}
\newcommand{\sect}[1]{Sec.~\ref{#1}}

\newcommand{\la}{\langle}
\newcommand{\ra}{\rangle}

\begin{document}

\title{Single-parameter aging in the weakly nonlinear limit} 
\author{Saeed Mehri}
\author{Lorenzo Costigliola}
\author{Jeppe C. Dyre}\email{dyre@ruc.dk}
\affiliation{{\it Glass and Time}, IMFUFA, Department of Science and Environment, Roskilde University, P. O. Box 260, DK-4000 Roskilde, Denmark} 
\date{\today}

\begin{abstract}
Physical aging deals with slow property changes over time caused by molecular rearrangements. This is relevant for non-crystalline materials like polymers and inorganic glasses, both in production and during subsequent use. The Narayanaswamy theory from 1971 describes physical aging -- an inherently nonlinear phenomenon -- in terms of a linear convolution integral over the so-called material time $\xi$. The resulting ``Tool-Narayanaswamy (TN) formalism'' is generally recognized to provide an excellent description of physical aging for small, but still highly nonlinear temperature variations. The simplest version of the TN formalism is single-parameter aging according to which the clock rate $d\xi/dt$ is an exponential function of the property monitored [T. Hecksher \textit{et al.}, J. Chem. Phys. \textbf{142}, 241103 (2015)]. For temperature jumps starting from thermal equilibrium, this leads to a first-order differential equation for property monitored, involving a system-specific function. The present paper shows analytically that the solution to this equation to first order in the temperature variation has a universal expression in terms of the zeroth-order solution, $R_0(t)$. Numerical data for a binary Lennard-Jones glass former probing the potential energy confirm that, in the weakly nonlinear limit, the theory predicts aging correctly from $R_0(t)$ (which by the fluctuation-dissipation theorem is the normalized equilibrium potential-energy time-autocorrelation function).
\end{abstract}

\maketitle

\section{Introduction}\label{intro}

The properties of non-crystalline materials like polymers and inorganic glasses change slightly over time. In many cases this aging is so slow that it cannot be observed, but sometimes it results in unwanted degradation of material properties. When aging is exclusively due to molecular rearrangements with no chemical reactions involved, one speaks about physical aging \cite{maz77,str78,scherer,hod95,cha02,pri09,gra12,can13,dyr13b,mic16,mck17,roth,rut17,mauro}. The present-day understanding of physical aging is based on the century-old observation \cite{sim31} that any glass is in an out-of-equilibrium state and, as a consequence, relaxes continuously toward the equilibrium state. 

During physical aging the system's volume decreases slightly. This reflects the fact that the equilibrium metastable liquid is denser than the glass at the same temperature. Likewise, the enthalpy decreases during aging. Both effects are extremely difficult to observe because they are minute and take place over a very long time. Defining a glass as any non-equilibrium state of a liquid resulting from a thermodynamic perturbation, a good way of studying physical aging is the following: First, equilibrate the glass-forming liquid at some ``annealing'' temperature just below the calorimetric glass-transition temperature. Depending on the viscosity of the liquid, this may take long time -- experiments often study liquids at temperatures at which the equilibrium relaxation time is hours or days \cite{ols98,hec10,hec15,hec19,roe19,rie22}. If the equilibrium relaxation time is one day, annealing the sample for a week ensures virtually complete thermal equilibrium. Once the sample has been equilibrated, temperature is changed rapidly to a new value and kept there for long enough time to allow for monitoring virtually the entire equilibration process. This defines an ``ideal aging experiment'' \cite{kol05,hec10}. Such an experiment requires the ability to monitor some quantity accurately and continuously as a function of time, excellent temperature control, and the ability to change temperature rapidly \cite{hec10}. Aging may be probed by measuring any property that can be monitored precisely, e.g., the electrical capacitance at a particular frequency \cite{sch91,leh98,lun05,ric15,rie22}; in conjunction with a Peltier-element-based fast and accurate temperature control this is our favorite method in Roskilde \cite{rie22}. Other quantities that have been monitored during physical aging include, e.g., density \cite{kov63,spi66}, enthalpy \cite{nar71,moy76}, Young’s modulus \cite{che78}, gas permeability \cite{hua04}, high-frequency mechanical moduli \cite{ols98,dil04}, dc conductivity \cite{str78}, X-ray photon correlation spectroscopy \cite{rut12}, and nonlinear dielectric susceptibility \cite{bru12}.

The present paper develops the theory of aging by studying the so-called single-parameter aging framework, which is the simplest realization of the concept of a \textit{material time} controlling aging in the Tool-Narayanaswamy (TN) formalism \cite{scherer}. An important prediction of the TN formalism is that if the aging rate is known as a function of the property monitored, knowledge of the linear limit of physical aging, e.g., following an infinitesimal temperature jump, is enough to quantitatively determine the aging resulting from any time-dependent temperature variation. According to the fluctuation-dissipation (FD) theorem any linear-response property is determined by thermal-equilibrium fluctuations quantified in terms of a time-autocorrelation function. The prospect for future experimental investigations is that one can make quantitative predictions of aging from a knowledge of equilibrium fluctuations.

Single-parameter aging results in a first-order differential equation for the normalized relaxation function following a temperature jump \cite{hec15}. This equation involves an \textit{a priori} unknown, system-specific function that determines the linear limit of aging. In order to predict aging, however, it is enough to know the linear-limit relaxation function that, by the FD theorem, is the relevant equilibrium time-autocorrelation function. This paper derives an explicit expression for the weakly nonlinear limit of aging based on the relevant equilibrium time-autocorrelation function. After developing the theory in \sect{TN}, we give an example of how to calculate a specific quantity similar to the fragility of glass science in \sect{perturb1}; this section can be skipped in a first reading of the paper. The general first-order solution to single-parameter aging following a temperature jump is derived in \sect{perturb2}. The validity of the formalism is illustrated in \sect{KA} by results from computer simulations of a binary Lennard-Jones system; the final section gives a brief discussion.

\section{The TN formalism and single-parameter aging}\label{TN}

The quantity probed during aging is denoted by $\chi(t)$. Following a temperature jump at $t=0$, $\chi(t)$ gradually approaches its equilibrium value $\chi_{\rm eq}$ at the new temperature $T_0$. We define the normalized relaxation function $R(t)$ by

\be\label{R_t}
R(t)\,\equiv\,\frac{\chi (t) - \chi_{\rm eq}}{\chi (0) - \chi_{\rm eq}}\,.
\ee
By definition, $R(t)$ is unity at $t=0$ and approaches zero as $t\to\infty$, i.e., for both temperature up and down jumps $R(t)$ is a decreasing function of time. In practice, in both experiments and simulations there is always a rapid initial change of $\chi(t)$ immediately after $t=0$ deriving from $\chi$'s dependence on the fast, vibrational degrees of freedom and/or one or more fast relaxation processes decoupled from the main ($\alpha$) relaxation. For this reason, workers in the field often normalize the relaxation function by defining $R(t)$ to be unity after the initial rapid change of $\chi(t)$. That is different from what is done in \eq{R_t}, which is our preference because it avoids introducing the extra parameters that comes from estimating the value of the short-time ``plateau'' of $\chi(t)$.

The TN \textit{material time} is denoted by $\xi$. This quantity, which may be thought of as the time measured on a clock with a clock rate $\gamma(t)$ that changes as the material ages, is related to the clock rate as follows  

\be\label{xi}
d\xi = \gamma(t)dt\,.
\ee
According to the TN formalism, the material time $\xi=\xi(t)$ controls the physical aging in such a way that the variation of $\chi(\xi)$, denoted by

\be
\Delta\chi(\xi)\equiv\chi (\xi)-\chi_{\rm eq} \,,
\ee
is a \textit{linear} convolution integral over the temperature variation history $T(\xi)-T_0$ \cite{nar71,scherer}.

Single-parameter aging (SPA) is the simplest version of the TN formalism \cite{hec15}. SPA assumes that the clock rate $\gamma(t)$ is an exponential function of the monitored property, i.e.,

\be\label{eq2a}
\gamma(t)\,=\,\geqa\, \exp\left(\dfrac{\Delta\chi(t)}{\chi_{0}}\right)\,.
\ee
Here $\geqa$ is the equilibrium relaxation rate at $T_0$ and $\chi_0$ is a constant with the same dimension as $\chi$. In conjunction with the TN prediction that physical aging is linear in the temperature variation when formulated in terms of the material time, SPA may be applied to any relatively small (continuous or discontinuous) temperature variation around $T_0$, not just to the discontinuous temperature jumps to which the below discussion is limited. Since $\Delta\chi(t)=\Delta\chi(0)R(t)$ by the definition of $R(t)$, \eq{eq2a} may be rewritten

\be\label{eq2}
\gamma(t)\,=\,\geqa \exp\left(\dfrac{\Delta\chi(0)}{\chi_{0}}R(t)\right)\,.
\ee

For temperature jumps the TN fundamental result is that \cite{nar71,scherer}

\be\label{eq6a}
R(t)=\Phi(\xi)\,
\ee
in which the function $\Phi(\xi)$ is the same for all temperature jumps of a given system. In view of the nonlinearity of physical aging, this is a highly nontrivial prediction. Keeping in mind the definition of $\gamma(t)$ (\eq{xi}), \eq{eq6a} implies $\dot{R}(t)=\Phi^\prime(\xi)\gamma(t)$. Since according to \eq{eq6a} $\xi$ is the same function of $R$ for all jumps, defining $F(R)\equiv - \Phi^\prime(\xi(R))$ leads to the ``jump differential equation''

\be\label{eq3}
\dot{R}(t) 
\,=\,-F(R)\gamma(t)
\,=\,-\geqa\,F(R)\, \exp\left(\dfrac{\Delta\chi(0)}{\chi_{0}}R(t)\right)\,
\ee
in which $F(R)$ is the same for all jumps of a given system. The negative sign in the definition of $F(R)$ is introduced in order to make $F(R)$ positive.

\Eq{eq3} has been confirmed in experiments on a silicone oil and several organic liquids \cite{hec15,roe19,rie22} aged to equilibrium just below their calorimetric glass transition temperature. Even though the largest temperature jumps studied were just a few percent, this is enough to exhibit a strongly nonlinear response with more than one decade of relaxation-time variation. One experimental test of \eq{eq3} involved rewriting it as  \cite{hec15,roe19}

\be\label{eq3a}
-\frac{\dot{R}(t)}{\geqa}\,\exp\left(-\dfrac{\Delta\chi(0)}{\chi_{0}}R(t)\right)
\,=\,F(R)\,
\ee
and showing that the left-hand side is the same function of $R$ for different jumps. A second test confirmed the consequence of \eq{eq3} that $R(t)$ for an arbitrary jump may be predicted from the data for a single jump \cite{hec15,roe19,rie22}.

This paper develops the SPA formalism based on \eq{eq3} that for simplicity is rewritten by adopting the unit system in which $\gamma_{\rm eq}=1$ at the reference temperature $T_0$:

\be\label{eq3b}
\dot{R} 
\,=\,-\,F(R)\, e^{\Lambda R}\,
\ee
with

\be\label{Lambda}
\Lambda\,\equiv\,\frac{\Delta\chi(0)}{\chi_{0}}\,.
\ee

\section{Calculation of a generalized fragility} \label{perturb1}

Each value of $\Lambda$ leads to a unique solution denoted by $R(t,\Lambda)$ of the jump differential equation \eq{eq3b} with the initial condition $R(0,\Lambda)=1$. As a first illustration of how perturbation theory may be applied when $|\Lambda|\ll 1$, we determined the $\Lambda$ dependence of the average relaxation time defined by

\be\label{tau_def}
\tau(\Lambda)
\,\equiv\,\int_0^\infty R(t,\Lambda)dt\,.
\ee
From $\tau(\Lambda)$ a fragility-like \cite{ang95} parameter $m_a$ (subscript ``a'' for aging) may be defined by 

\be\label{tau_var}
m_a
\,\equiv\,\left.-\, \frac{d}{d\Lambda}\ln\tau\right|_{\Lambda=0}\,.
\ee
The minus ensures that $m_a>0$ because $\Lambda>0$ from \eq{eq3b} leads to a faster than equilibrium relaxation. 

We proceed to derive the following expression in which $R_0(t)\equiv R(t,\Lambda=0)$

\be\label{ma_result}
m_a
\,=\,\frac{\int_0^\infty R_0^2(t)dt}{\int_0^\infty R_0(t)dt}\,.
\ee
Note that whenever $0<R_0(t)<1$, which is usually the case \cite{hec15}, one has $m_a<1$. Note also that since $\geqa(\Lambda)=\exp(\Lambda)$ from \eq{eq3b}, the equilibrium relaxation time $\teq(\Lambda)\equiv 1/\geqa(\Lambda)$ obeys $d\ln\teq=-d\Lambda$. Using this one can transform \eq{ma_result} into an expression for how the relative change of $\tau(\Lambda)$ from its value at $T_0$ depends on the relative change of the equilibrium relaxation time between the two temperatures involved in the jump, i.e.,

\be\label{ma_result2}
\left.\, \frac{d\ln\tau}{d\ln\teq}\right|_{T=T_0}
\,=\,\frac{\int_0^\infty R_0^2(t)dt}{\int_0^\infty R_0(t)dt}
\,=\,m_a\,.
\ee
The fact that $m_a<1$ is now intuitively obvious since the graph of $R(t)$ obviously falls between the equilibrium relaxation function graphs at the two temperatures.

To derive \eq{ma_result}, note that \eq{eq3b} implies $dt=-\exp(-\Lambda R)\,dR/F(R)$. Thus  

\be
\tau(\Lambda)
\,=\,-\int_1^0\frac{R\,e^{-\Lambda R}}{F(R)}dR
\,=\,\int_0^1\frac{R\,e^{-\Lambda R}}{F(R)}dR\,.
\ee
From this we get

\be
\tau(\Lambda=0)
\,=\,\int_0^1\frac{R}{F(R)}\,dR\,.
\ee
and

\be
\left.\frac{d\tau}{d\Lambda}\right|_{\Lambda=0}
\,=\,-\int_0^1\frac{R^2}{F(R)}\,dR\,.
\ee
By substituting $R=R_0$ into both integrals and switching back to time as the integration variable one finds

\be\label{slutresut}
m_a
\,=\,\frac{\int_0^1\frac{R_0^2}{F(R_0)}\,dR_0}{\int_0^1\frac{R_0}{F(R_0)}\,dR_0}
\,=\,\frac{\int_0^\infty R^2_0(t)\,dt} {\int_0^\infty R_0(t)\,dt}\,.
\ee
An alternative proof of \eq{ma_result} makes use of an integral criterion derived in Ref. \onlinecite{hec15} (Appendix). 

For the calculation of $m_a$ from experimental or computer simulation data on $R_0(t)$ one proceeds as follows. Given a sequence of times $(\Delta t, 2\Delta t,3\Delta t,...,n\Delta t)$ at which the equilibrium normalized relaxation function $(R_{0,1},R_{0,2},R_{0,3},...,R_{0,n})$ is known, we have

\be\label{ma_eq2}
m_a
\,=\,\,\frac{\sum_{j=1}^n R_{0,j}^2}{\sum_{j=1}^n R_{0,j}}\,.
\ee

As an illustration of the above we consider the case in which the linear-limit relaxation function is a stretched exponential with exponent $\beta$ (for simplicity, a dimensionless time is used below),

\be\label{strexp1}
R_0(t)
\,=\,e^{- t^\beta}\,.
\ee
Defining the function 
\be\label{strexp2}
f(\alpha,\beta)
\,=\,\int_0^\infty e^{-\alpha t^\beta}dt\,,
\ee
we have $m_a=f(2,\beta)/f(1,\beta)$. Since

\be
f(2,\beta)
\,=\,\int_0^\infty e^{-2t^\beta}dt
\,=\,2^{-1/\beta}\,\int_0^\infty e^{-2t^\beta}d(2^{1/\beta}t)
\,=\,2^{-1/\beta}\,f(1,\beta)\,,
\ee
we get

\be
m_a
\,=\,2^{-1/\beta}\,.
\ee
If the normalized relaxation function in the experimental time window is described by a stretched exponential with short-time plateau $C<1$, i.e., by the function $C\exp(-t^\beta)$, one finds

\be
m_a
\,=\,C\,2^{-1/\beta}\,.
\ee

\section{Solving the jump differential equation to first order in the temperature change $\Delta T$}\label{perturb2}

To find the solution, $R(t,\Lambda)$, of the jump differential equation in first-order perturbation theory we proceed as follows. A first-order expansion of $R(t)$,

\be\label{Rt}
R(t)
\,=\,R_0(t)+\Lambda R_1(t)\,,
\ee
is substituted into \eq{eq3b} where $R_0(t)$ is the normalized relaxation function corresponding to an infinitesimal jump, i.e., to the linear limit of aging (this function is discussed below in \sect{sec:fd}). To first order in $\Lambda$ one has $F(R)=F(R_0)+ F'(R_0)\Lambda R_1$ and $\exp(\Lambda R)=1+\Lambda R=1+\Lambda R_0$. This results in

\be\label{first_order_eq2}
\dot R_0 + \Lambda \dot R_1
\,=\,-\big(F(R_0)+F'(R_0)\Lambda R_1\big)\big(1+\Lambda R_0\big)\,,
\ee
which leads to the following zeroth- and first-order equations:

\begin{eqnarray}
\dot R_0 &=& -F(R_0)\label{array1}\\
\dot R_1 &=& -F(R_0)R_0-F'(R_0)R_1\label{array2}\,.
\end{eqnarray}
Due to the zero-time normalization of both $R(t)$ and $R_0(t)$, the initial condition of $R_1$ is $R_1(0)=0$. For $t>0$ one has $R_1(t)<0$ because $\Lambda>0$ as mentioned implies a faster relaxation toward equilibrium, i.e., $R(t)<R_0(t)$. Consequently, since $R_1(0)=R_1(t\to\infty)=0$, the function $R_1(t)$ is non-monotonous. 

The solution to \eq{array2} obeying the initial condition $R_1(0)=0$ is

\be\label{R1_sol1}
R_1(t)
\,=\,\,\dot R_0(t)\int_{0}^tR_0(t')dt'\,.
\ee
To derive this we proceed as follows. First, we note that the inverse of $R(t,\Lambda)$ is given by

\be\label{sol}
t(R,\Lambda)
\,=\,-\int_1^R e^{-\Lambda R'}\frac{dR'}{F(R')}\,,
\ee
which follows by rewriting \eq{eq3b} as $dt=-\exp(-\Lambda R)\,dR/F(R)$ and integrating. Next we note that because $R(t,0)=R_0(t)$,

\be\label{R1}
R_1(t)
\,=\,\left(\frac{\partial R}{\partial\Lambda}\right)_t
\,=\,-\frac{\left(\frac{\partial t}{\partial\Lambda}\right)_R}
{\left(\frac{\partial t}{\partial R}\right)_\Lambda}\,
\ee
in which it here and henceforth is understood that all functions are evaluated at $\Lambda=0$, implying that one should put $R=R_0$ in the final evaluations. Noting that $dR_0/F(R_0)=-dt$, using \eq{sol} the numerator is evaluated as follows

\be\label{numer}
\left(\frac{\partial t}{\partial\Lambda}\right)_R
\,=\,\int_1^R R'\frac{dR'}{F(R')}
\,=\,-\int_0^t R_0\,dt'\,.
\ee
For $\Lambda=0$ the denominator of \eq{R1} is given by

\be\label{denom}
\left(\frac{\partial t}{\partial R}\right)_\Lambda
\,=\,\frac{1}{\dot{R}_0(t)}\,.
\ee
Combining these results one arrives at \eq{R1_sol1}. To confirm that \eq{R1_sol1} indeed solves \eq{array2}, one differentiates:

\be\label{sol1}
\dot R_1(t)
\,=\,\ddot R_0(t) \int_{0}^tR_0(t')dt'\,+\,\dot R_0(t)R_0(t)\,.
\ee
Since $\ddot R_0=-F'(R_0)\dot R_0$ by \eq{array1}, we see that $\dot R_1=-F'(R_0)R_1-F(R_0)R_0$ as required.

As an illustration, we show how \eq{R1_sol1} leads to \eq{ma_result}. From \eq{tau_def}, \eq{tau_var}, and \eq{Rt} one easily derives 

\be\label{ma_eq}
m_a
\,=\,-\,\frac{\int_{0}^{\infty}R_1(t)dt}{\int_{0}^{\infty}R_0(t)dt}\,.
\ee
This is simplified by performing a partial integration:

\begin{eqnarray}\label{eq11}
-\int_0^\infty R_1(t)dt
\,&=&\,-\int_0^\infty \dot R_0(t)\left(\int_0^t R_0(t')dt'\right)dt\nonumber\\
\,&=&\,-\left[ R_0(t)\left(\int_0^t R_0(t')dt'\right)\right]_0^\infty
\,+\,\int_0^\infty R_0^2(t)dt\nonumber\\
\,&=&\,\int_0^\infty R_0^2(t)dt\,.
\end{eqnarray}
We thus arrive at \eq{ma_result}.

\section{Numerical results for a binary Lennard-Jones model}\label{KA}

\begin{figure}[!h]
    \includegraphics[width=0.45\textwidth]{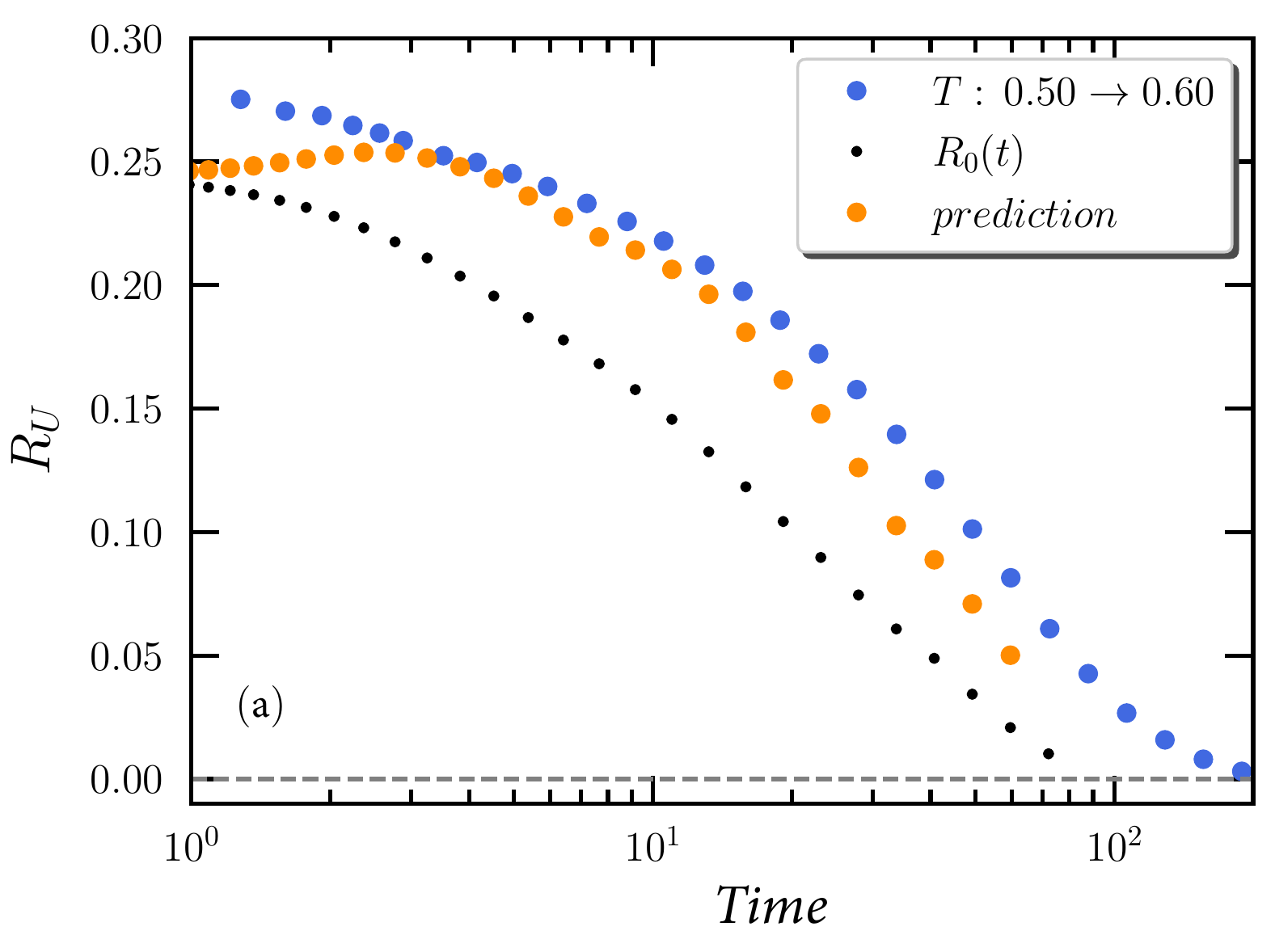}
    \includegraphics[width=0.45\textwidth]{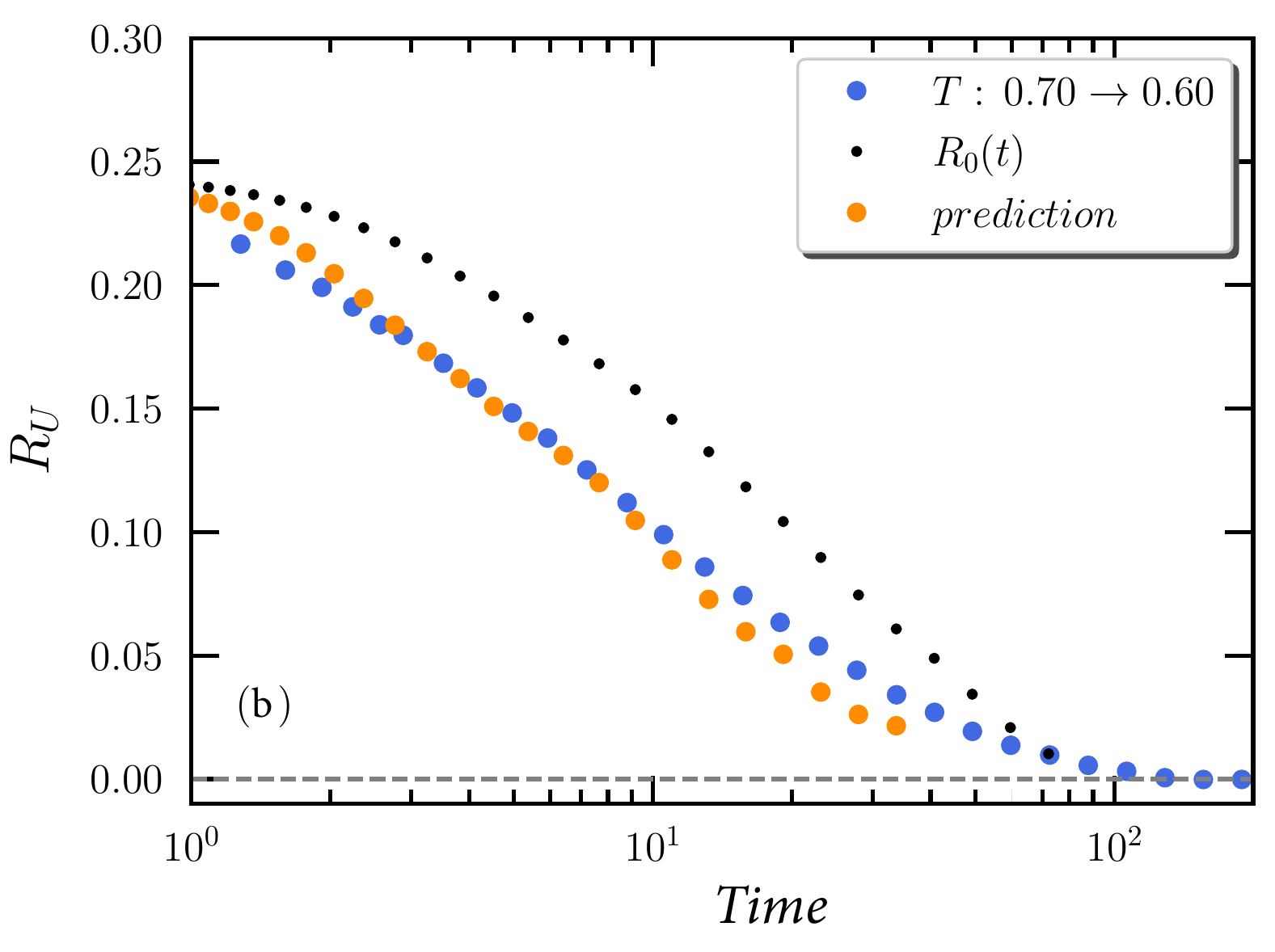}
    \includegraphics[width=0.45\textwidth]{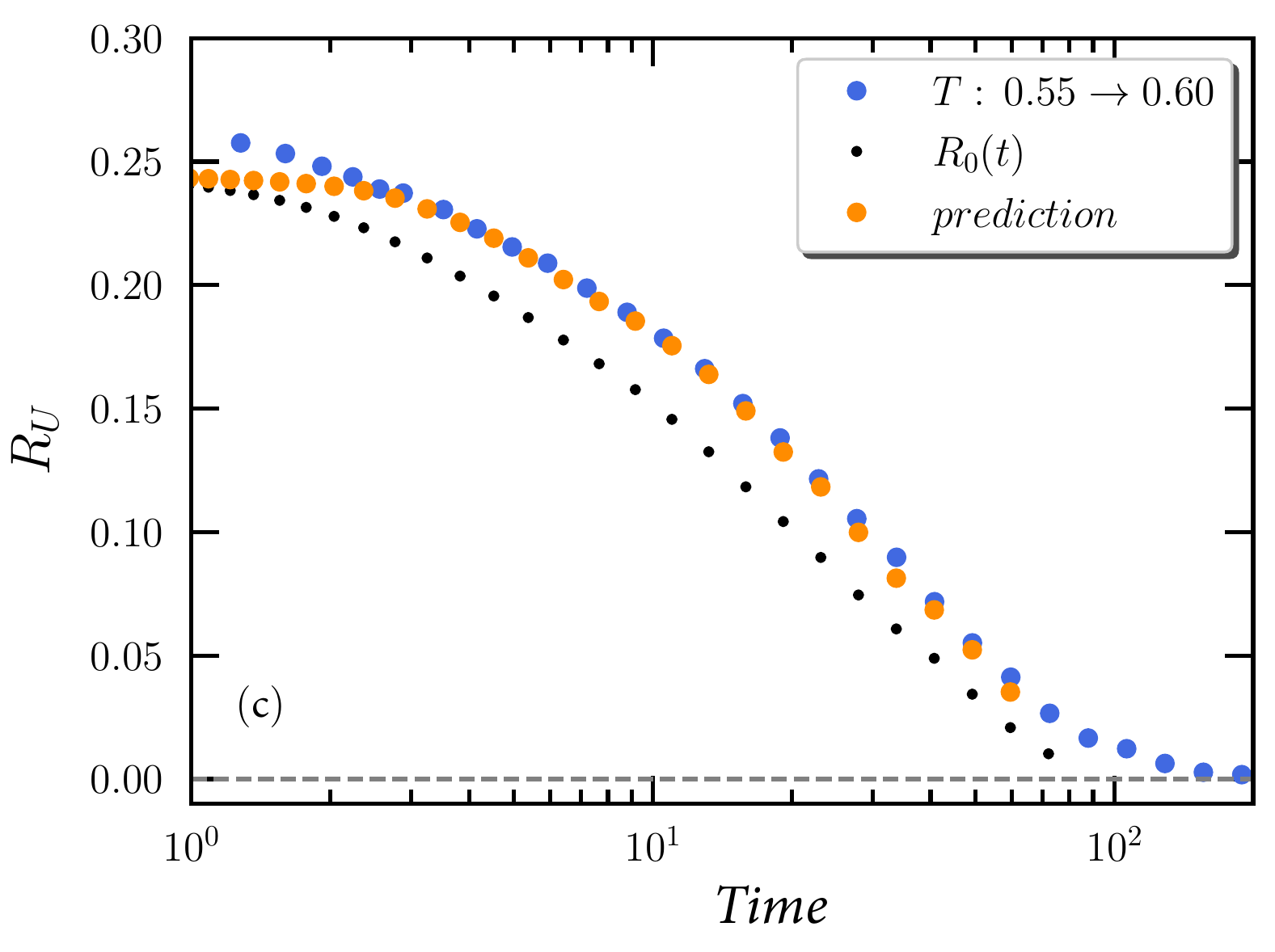}
    \includegraphics[width=0.45\textwidth]{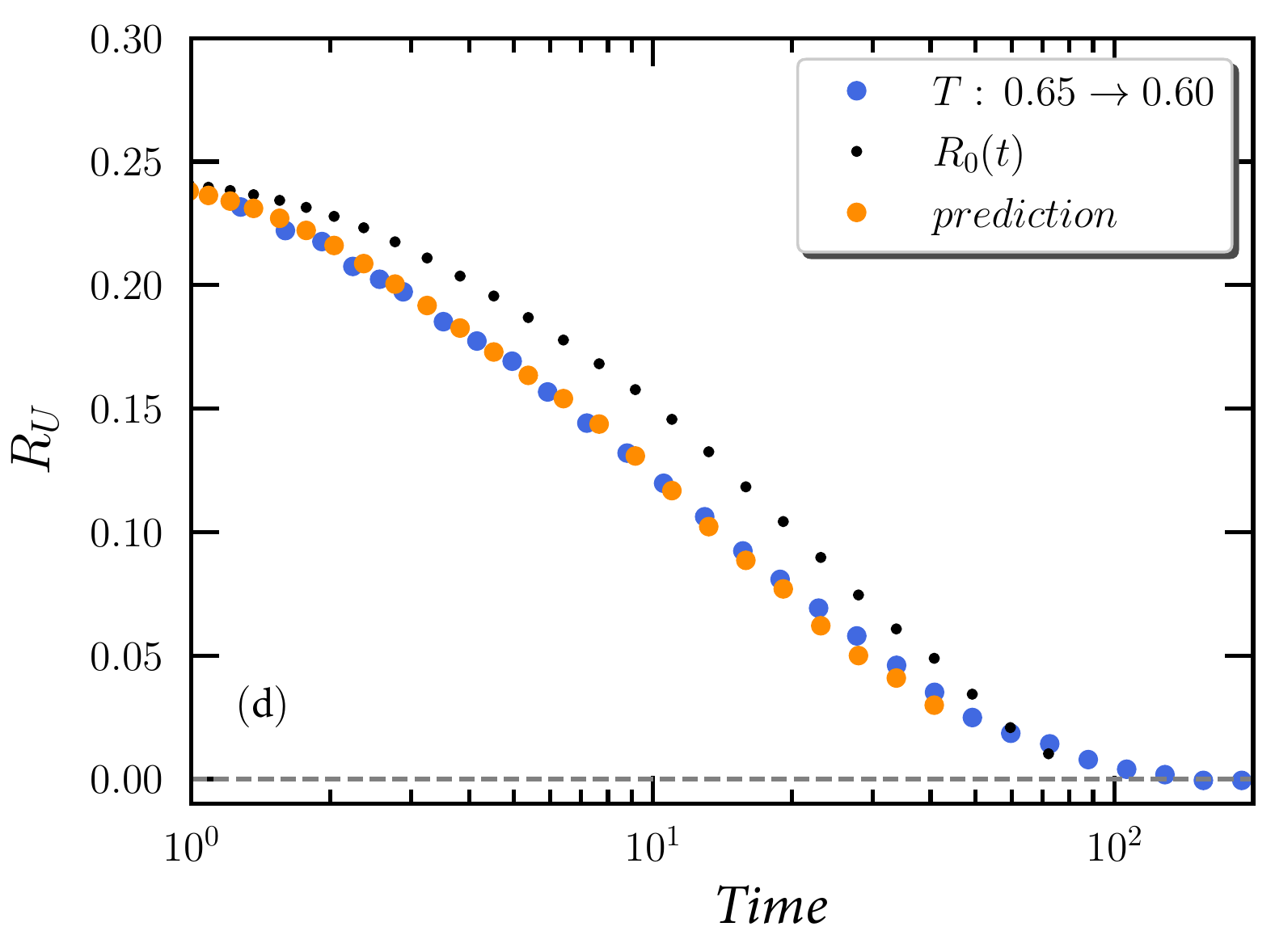}    
    \includegraphics[width=0.45\textwidth]{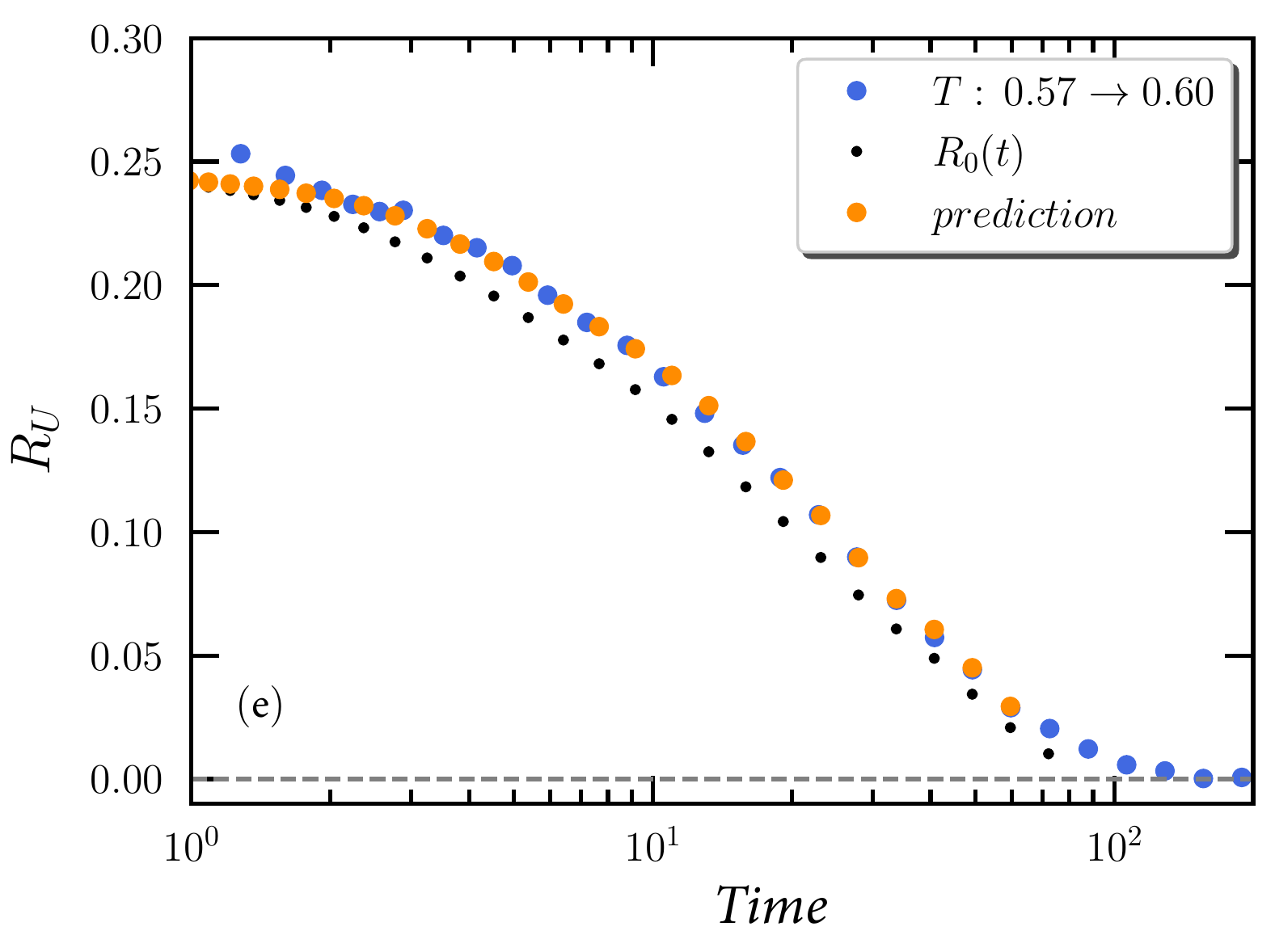}
    \includegraphics[width=0.45\textwidth]{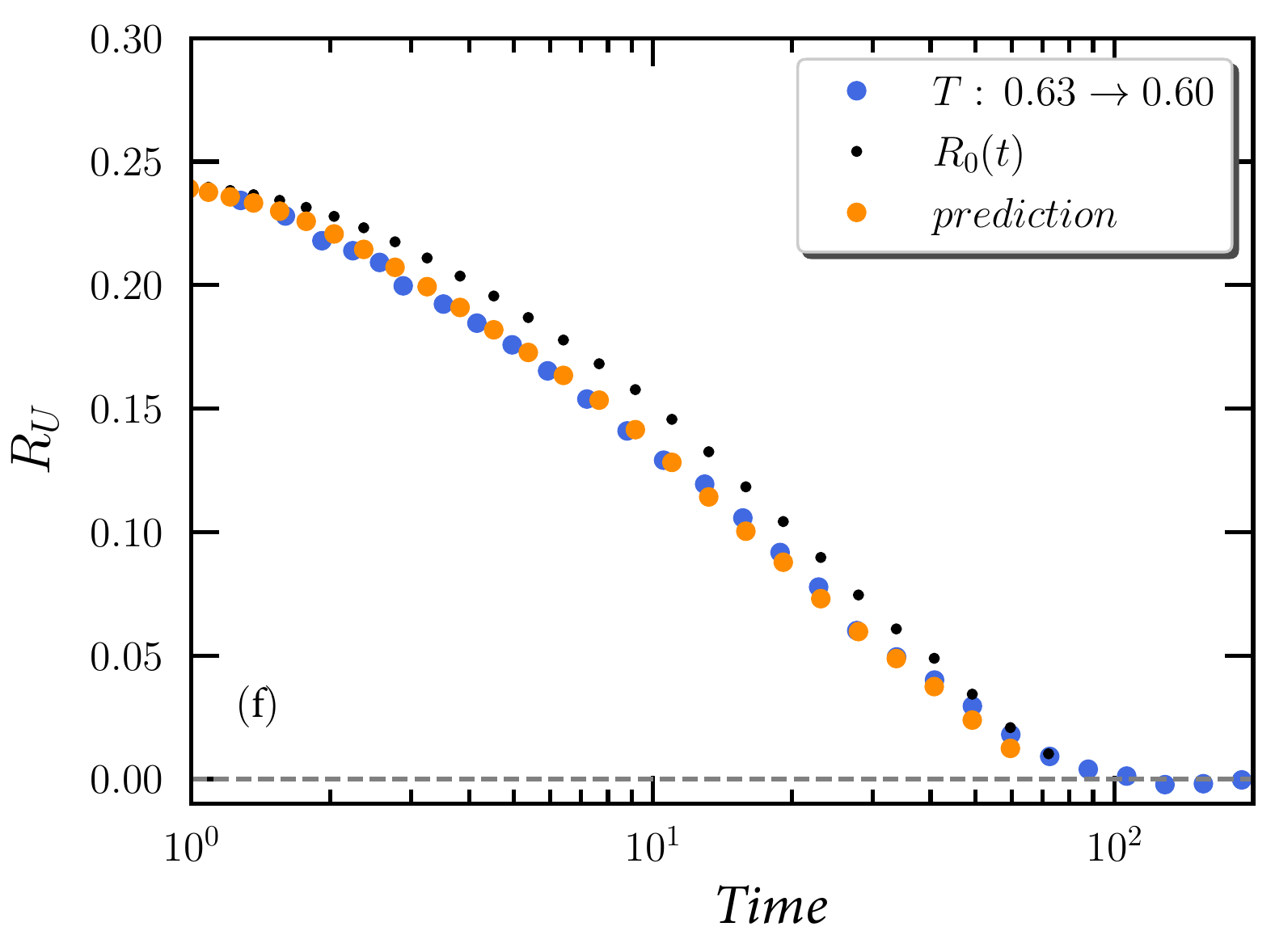}
    \caption{Results from computer simulations of the Kob-Andersen binary Lennard-Jones model. The figures show the normalized relaxation function $R(t)$ (\eq{Rt} and \eq{R1_sol1}) defined from the potential energy $U$ after a temperature jump at $t=0$ starting from a state of thermal equilibrium (blue filled circles). 
    (a) and (b) show results for magnitude 0.10 temperature up and down jumps to the reference temperature $T_0=0.60$;
    (c) and (d) show results for magnitude 0.05 temperature up and down jumps to $T_0=0.60$;
    (e) and (f) show results for magnitude 0.03 temperature up and down jumps to $T_0=0.60$.
    The orange filled circles are the first-order predictions of the jump differential equation \eq{eq3b} (given in \eq{Rt} in which $R_0(t)$ is the normalized equilibrium potential-energy time-autocorrelation function at $T_0=0.60$, $R_1(t)$ is given by \eq{R1_sol1}, and $\Lambda$ is given by \eq{Leq}). For reference, $R_0(t)$ in all figures is plotted as small black filled circles.}   \label{Fig1}
\end{figure}

\subsection{The relevant fluctuation-dissipation theorem}\label{sec:fd}

When the externally controlled variable is temperature itself, a slightly modified derivation of the FD theorem is required \cite{nie96}. In the end, however, the result looks much like in the standard FD case: If $\Delta\beta(t)$ is the variation of $\beta\equiv 1/k_BT$ from its equilibrium value at the reference temperature, $\delta\beta(t)\equiv\beta(t+dt)-\beta(t)$, and sharp brackets denote standard canonical averages, the variation of the potential energy is given \cite{nie96} by 

\be\label{fd_theorem}
\DU(t)
\,=\, -\DUto\,\Delta\beta(t)\,+\,\int_{-\infty}^t \la\DU(0)\DU(t-t')\ra\,\delta\beta(t')\,.
\ee
Following a small inverse-temperature step of magnitude $\Delta\beta$, $\DU(t)\to -\DUto\,\Delta\beta$ for $t\to\infty$. Since $\Delta\beta=-\Delta T_0/k_BT_0^2$, this leads to the well-known Einstein expression for the specific heat, $c=\DUto/k_BT_0^2$. 

\Eq{fd_theorem} implies that the response to a jump at $t=0$ for $t>0$ is given by  $\DU(t)=\left[-\DUto+\la\DU(0)\DU(t)\ra\right]\Delta\beta$ (note that right after the temperature step one has $\DU(t)\cong 0$ because of continuity). Therefore, the linear-response normalized relaxation function, $R_0(t)$, is given by 

\be\label{RU_exp}
R_0(t)
\,=\,\frac{\DU(t)-\DU(t=\infty)}{\DU(0)-\DU(t=\infty)}
\,=\,\frac{\la\DU(0)\DU(t)\ra}{\DUto}\,.
\ee

\subsection{Simulation results}

We simulated the well-known Kob-Andersen binary Lennard-Jones (KA) 80/20 mixture of A and B particles \cite{ka1} with the standard Nose-Hoover thermostat \cite{nose} by means of the GPU-software RUMD \cite{RUMD}. The pair potentials of the KA system are Lennard-Jones potentials, i.e., $v_{ij}(r) = \varepsilon_{ij}[(\sigma_{ij}/r)^{12} - (\sigma_{ij}/r)^{6}]$ ($i,j=A,B$) with the following parameters: $\sigma_{AA} = 1$, $\sigma_{AB} = 0.80$, $\sigma_{BB} = 0.88$, $\varepsilon_{AA} = 1$, $\varepsilon_{AB} = 1.5$, $\varepsilon_{BB} = 0.5$. All masses are set to unity. A system of $N$ = 8000 particles was simulated. In the units based on $\sigma_{AA}$ and $\varepsilon_{AA}$, the time step was $\Delta t$ = 0.0025 and the thermostat relaxation time was $0.2$. The potentials were cut and shifted at $r_{\rm c}$ = 2.5$\sigma_{ij}$.

The potential-energy time-autocorrelation function appearing in \eq{RU_exp} was calculated at the reference temperature $T_0=0.60$ as follows. First, $10^7$ time steps were taken for equilibration, which was confirmed from two consecutive runs comparing the self-part of the intermediate scattering function. After that a run of $5\times 10^{6}$ time steps was carried out dumping the potential energy every 32 time steps. The potential-energy time-autocorrelation function was calculated using Fast Fourier Transform as implemented in RUMD \cite{tildesley}. 

In SPA the constant $\Lambda$ of \eq{eq3b} is assumed to be proportional to the change in the monitored property, \textit{in casu} $\Delta U$. $\Lambda$ was determined using the integral criterion of Ref. \cite{hec15}, which considers two jumps to the same temperature, an up and a down jump. For this we used the jumps from the temperatures 0.55 and from 0.65 to $T_0=0.60$ \cite{rie22}, leading to

\be\label{Leq}
\Lambda
\,=\,\frac{\Delta U}{0.0404}\,.
\ee
Here $\Delta U$ is the equilibrium potential energy at the starting temperature minus the corresponding quantity at the final temperature (following the tradition in the field). The $\Lambda$ of \eq{Leq} was used for all predictions.

Temperature-jump simulations were carried out as follows. First, $5 \times 10^8$ time steps were taken to ensure equilibration at the given starting temperature. A total of 1000 configurations were generated from a subsequent $5 \times 10^{8}$ simulation by dumping configurations every $2^{19}$ time step. This ensures that the configurations are statistically independent at the lowest temperature studied ($T$ = 0.50). For each of the 1000 configurations an aging simulation of $10^{6}$ time steps was performed and the potential energy was dumped every eighth time steps. The curves shown in \fig{Fig1} represent averages over these 1000 aging simulations.

The averages were smoothed using a Gaussian function. Each point represents an average calculated over all data points using $R_{\rm avg} (t) = {\sum_{t'} R(t') \exp(-(t-t')^2/\sigma)}/{\sum_{t'} \exp(-(t-t')^2/\sigma)}$ in which $t'$ is the time-step number and $\sigma = 15,000$. In order to reduce the number of points shown in \fig{Fig1}, the data were divided into 24 bins per decade.

\Fig{Fig1} shows the simulation results (blue circles) for the normalized relaxation function of the potential energy for up and down jumps to $T_0=0.60$. The orange circles are the predictions of the first-order theory. In all figures the small filled black circles are the normalized equilibrium potential-energy time-autocorrelation function at $T_0=0.60$, which is the linear-limit normalized relaxation function $R_0(t)$ (\eq{RU_exp}). This function is faster than $R(t)$ for up jumps and slower for down jumps. This is expected since relaxation is initially slow for the up jumps to $T_0=0.60$ because the fictive temperature \cite{scherer} in this case is below 0.60, while the opposite happens for the down jumps to $T_0=0.60$. 

We see that the theory generally fits data well, even for the fairly large temperature jumps of magnitude 0.05. Deviations between prediction and simulations is observed for larger up jumps, though. A similar pattern has been observed in experiments but there the observed relaxation function is faster than predicted, not slower \cite{rie22}. In both cases these deviations serve to emphasize that SPA is not accurate for large jumps.

\begin{figure}[!h]
    \includegraphics[width=0.45\textwidth]{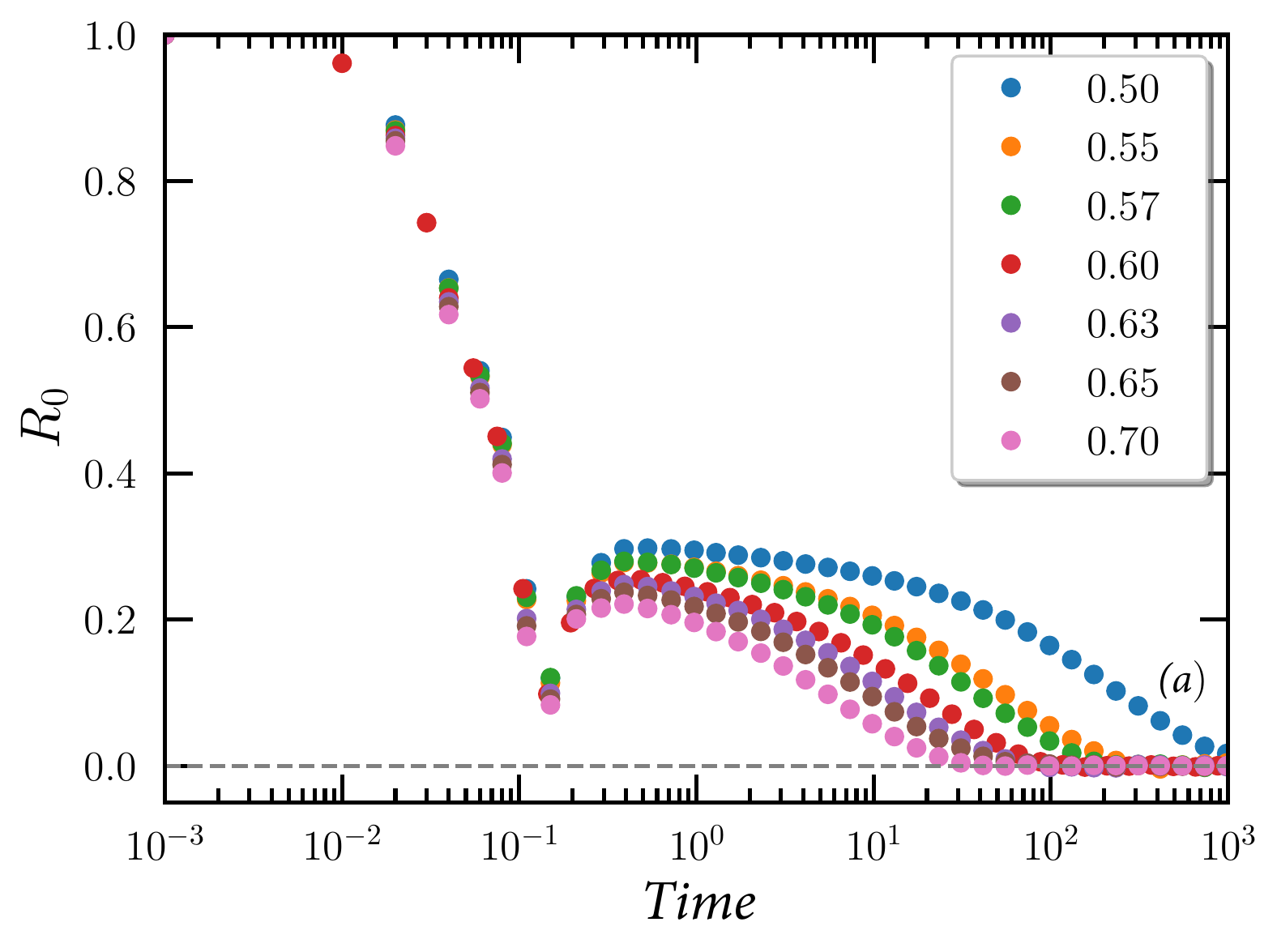}
    \includegraphics[width=0.45\textwidth]{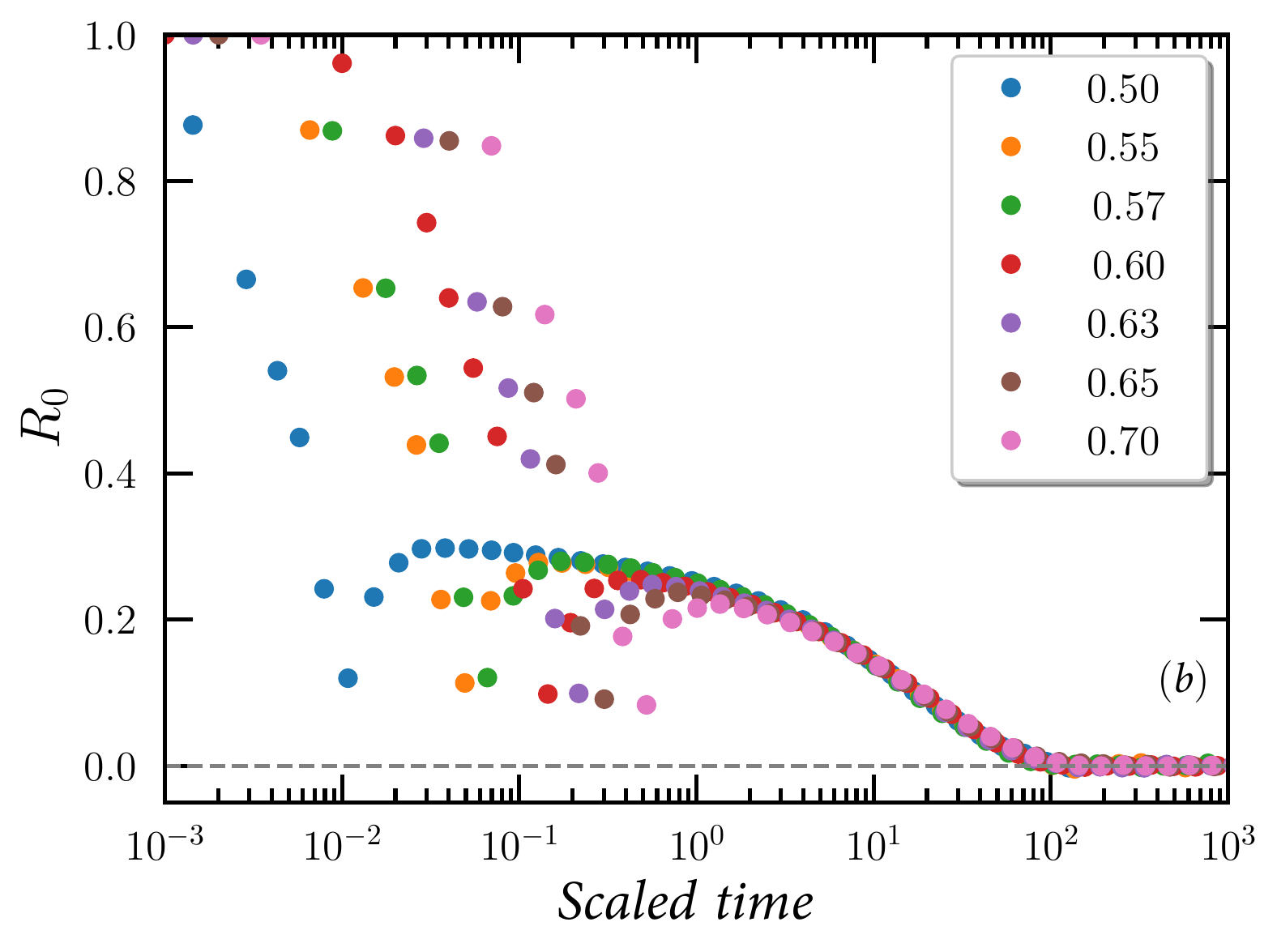}
    \caption{Test of time-temperature superposition for the normalized potential-energy time-autocorrelation function $R_0(t)$ at the temperatures indicated in the legends. 
    (a) shows the simulation data and (b) shows the same data empirically scaled on the time axis. We conclude that TTS applies except at the shortest times.}
    \label{Fig2}
\end{figure}

The TN formalism implies that the long-time decay of the normalized relaxation function for infinitesimal jumps to different temperatures are identical except for an overall scaling of the time, i.e., obeys time-temperature superposition (TTS) \cite{ols01}. In other words, TTS is a necessary condition for TN to apply and thus, in particular, for SPA to apply. We test TTS by plotting the normalized potential-energy time-autocorrelation functions $R_0(t)$ at temperatures ranging from 0.50 to 0.70 (\fig{Fig2}(a)) and scaling these on the time axis (\fig{Fig2}(b)). Except for the short-time signals that are not relevant to aging, we see that TTS indeed applies to a very good approximation.

\section{Summary and outlook}

We have solved the jump differential equation analytically to first order. The solution is \eq{Rt} in which $R_1(t)$ is given by \eq{R1_sol1}. The solution does not explicitly involve the function $F(R)$; indeed $R_1(t)$ has a universal expression in terms of the zeroth-order solution, $R_0(t)$. Since the latter by the fluctuation-dissipation theorem is an equilibrium time-autocorrelation function, our results imply that within the single-parameter aging scheme knowledge of equilibrium fluctuations is enough to predict aging. The expression for $R_1(t)$ relevant for the weakly nonlinear limit was confirmed by computer simulations of the Kob-Andersen binary Lennard-Jones glass former monitoring the aging of the potential energy following temperature jumps of varying magnitude.

For future developments of the TN single-parameter aging formalism it would be most interesting to monitor the equilibrium fluctuations in experiments in order to check whether aging is predicted correctly from these fluctuations. This is experimentally very challenging, but should not be impossible. It would also be interesting to monitor other quantities in simulations than the potential energy used here, although it should be mentioned that many quantities relax in a very similar way for the Kob-Andersen system \cite{meh21}.

\section*{Appendix}

\Eq{ma_result} is derived here from the integral criterion of Ref. \cite{hec15} that considers two jumps to the same temperature: an up and a down jump. For two small such jumps of same magnitude to the temperature $T_0$ denoted by $a$ and $b$, one has the two normalized relaxation functions

\begin{eqnarray}\label{Raeq}
R_a\,&=&\,R_0\,+\,\Lambda R_1\nonumber\\
R_b\,&=&\,R_0\,-\,\Lambda R_1\,.
\end{eqnarray}
The integral criterion \cite{hec15} is

\be\label{crit}
\int_0^\infty\left(e^{\Lambda_{ab}R_a}-1\right)dt\,+\,\int_0^\infty\left(e^{\Lambda_{ba}R_b}-1\right)dt
\,=\,0\,.
\ee
Here $\Lambda_{ab}=-\Lambda_{ba}$ is the difference in the value of $\Lambda$ jumping from above and below, implying that $\Lambda_{ab}=2\Lambda$ and $\Lambda_{ba}=-2\Lambda$. When \eq{Raeq} is substituted into \eq{crit}, we get

\be
\int_0^\infty\left(e^{2\Lambda (R_0+\Lambda R_1)}-1\right)dt\,+\,\int_0^\infty \left(e^{-2\Lambda (R_0-\Lambda R_1)}-1\right)dt
\,=\,0\,.
\ee
Expanding to second order in $\Lambda$ leads to

\be
\int_0^\infty \Big(2\Lambda (R_0+\Lambda R_1)+2\Lambda^2 R_0^2
+(-2\Lambda (R_0-\Lambda R_1))+2\Lambda^2 R_0^2\Big)dt
\,=\,0\,.
\ee
This reduces to

\be
\int_0^\infty \Big( 4\Lambda^2 R_1 + 4\Lambda^2 R_0^2\Big)dt
\,=\,0\,,
\ee
which implies \eq{eq11} and thereby \eq{ma_result}.

\end{document}